\documentclass[%
 reprint,
 amsmath,amssymb,
 aps, prl,
]{revtex4-1}

\usepackage{graphicx}
\usepackage{dcolumn}
\usepackage{bm}
\usepackage{mathtools}

\begin{document}


\title{Homoclinic chaos and its organization in a nonlinear optics model}
\author{Krishna Pusuluri}
\email{pusuluri.krishna@gmail.com}
\affiliation{Neuroscience Institute, Georgia State University, Petit Science Center, 100 Piedmont Av., Atlanta,
GA 30303, USA}
\author{Andrey Shilnikov}
\email{ashilnikov@gsu.edu}
\affiliation{Neuroscience Institute, and Department of Mathematics and Statistics, Georgia State University, Petit Science Center, 100 Piedmont Av., Atlanta, GA 30303, USA}
\date{\today}
\begin{abstract}

\begin{description}
\item[Abstract]
We developed a powerful computational approach to elaborate on onset mechanisms of deterministic chaos due to complex homoclinic bifurcations in diverse systems. Its core is the reduction of phase space dynamics to symbolic binary representations that lets one detect regions of simple and complex dynamics as well as fine organization structures of the latter in parameter space. Massively parallel simulations shorten the computational time to disclose highly detailed bifurcation diagrams to a few seconds.      
\end{description}
\end{abstract}

\maketitle

{\em New directions in science are launched by new tools much more often than by new concepts.}~F.Dyson \cite{dyson1998}. 
%
 Break-through discovery of deterministic chaos in [infra-red gas] lasers in nonlinear optics was pioneered and established both theoretically and experimental long time ago  \cite{haken1975,haken1985,Casperson1978,WeissandKing1982,Weissetal1985,WeissandBrock1986}. Recent developments in semiconductor lasers and nano-optics have stimulated newest advances in optical  synchronization and photonic integrated circuits for needs of cryptography \cite{pecora1990, ott1990, liu1995, colet1994, royR1998, uchida2008, argyris2008, ohtsubo2012, naruse2014}. 
 Nowadays, a real advance in deterministic nonlinear science stimulating the progress in cutting-edge engineering is hardly possible without significant deepening the knowledge (know-how) and beneficial usage of complex elements borrowed from dynamical systems theory, which in turn is hardly possible without development and incorporation of new mathematical and computational tools, including for parallel [GPU-based] platforms. \\
 \noindent
 In this letter we demonstrate how our newly developed toolkit, called ``Deterministic Chaos Prospector (DCP)'' lets one quickly and fully disclose and elaborate on the origins of complex chaotic dynamics in a 6D model of a resonant 3-level optically-pumped laser (OPL) \cite{harrison1,harrison2}. In addition to simple dynamics associated with stable equilibria and periodic orbits, it reveals a broad range bifurcation structures that are typical for many deterministic models from nonlinear optics and other applications~\cite{bss,xwbs,xbs,xwzs2015,pusuluri2017}. These   include  homoclinic orbits and heteroclinic connections between saddle equilibria, which are the key building blocks of deterministic chaos. Their bifurcation curves with characteristic spirals around T-points along with other codimension-2 points are the organizing centers that shape regions of complex and simplex  dynamics in parameter space of such systems. The detection of these bifurcations has long remained the state-of-the-art involving a meticulous  and time consuming parameter continuation technique to disclose a few  sparse elements of the otherwise rich and fine organization of the bifurcation set. We note that while the brute-force approach based on the evaluation of Lyapunov exponents can locate stability windows within regions of chaos \cite{gallas,bbss}, it fails to disclose essential structures that are imperative for understanding complex dynamics and its origin. We will also demonstrate how our approach exploiting the sensitivity of deterministic chaos and its symbolic representation  using binary sequences, along with the Lempel-Ziv complexity algorithms \cite{lz76}, can effectively reveal regions of complex, structurally unstable and simple stable dynamics in this and other systems.\\ 
\noindent
The 3-level optically pumped laser model~\cite{harrison1,harrison2} is given by  
\begin{eqnarray}
\dot{\beta} ~~&=& -\sigma \beta + 50 p_{23}, \nonumber  \\
\dot{p}_{21} &=& -p_{21} - \beta p_{31} + a D_{21}, \nonumber\\
\dot{p}_{23} &=& -p_{23} + \beta D_{23} - a p_{31}, \nonumber\\
\dot{p}_{31} &=& -p_{31} + \beta p_{21} + a p_{23}, \\
\dot{D}_{21} &=& -b(D_{21}-D_{21}^0)-4 a p_{21} - 2 \beta p_{23}, \nonumber\\
\dot{D}_{23} &=& -b(D_{23}-D_{23}^0)-2 a p_{21} - 4 \beta p_{23}, \nonumber
\end{eqnarray}
 with bifurcation parameters $a$, $\beta$, and $\sigma=\{1.5;\,10\}$, being the Rabi flopping quantities representing the electric field amplitudes at pump and emission frequencies, and the cavity loss parameter, resp.; $b$ is the ratio of population to polarization decay rates; $p_{ij}$'s represents the normalized density matrix elements corresponding to the  transitions between levels $i$ and $j$, while $D_{ij}$ is the population difference between the $i$-th and $j$-th levels. Note that Eqs.~(1) are $\mathbb{Z}_2$-symmetric under involution $(\beta,p_{21},p_{23},p_{31},D) \leftrightarrow (-\beta,p_{21},-p_{23},-p_{31},D)$, which is typical for Lorenz-line systems \cite{abs,xbs}. 
Depending on $(a,\beta)$-values, the laser model~(1) has either a single non-lasing steady state, $O$, or an extra pair of equilibria, $C^\pm$ (Fig.~1a), emerging as $0$ loses stability through a pitch-fork $PF$ bifurcation and becomes a saddle. All three steady states can independently undergo super-critical Andronov-Hopf ($AH$) bifurcations (curves labelled with $AH_0$ and $AH_{1,2}$ in the $(a,\,b)$-parameter plane in Fig.2) giving rise to stable periodic orbits (PO) in the phase space of the laser model. 
\begin{figure}[ht!]
     \begin{center}
        \includegraphics[width=\columnwidth]{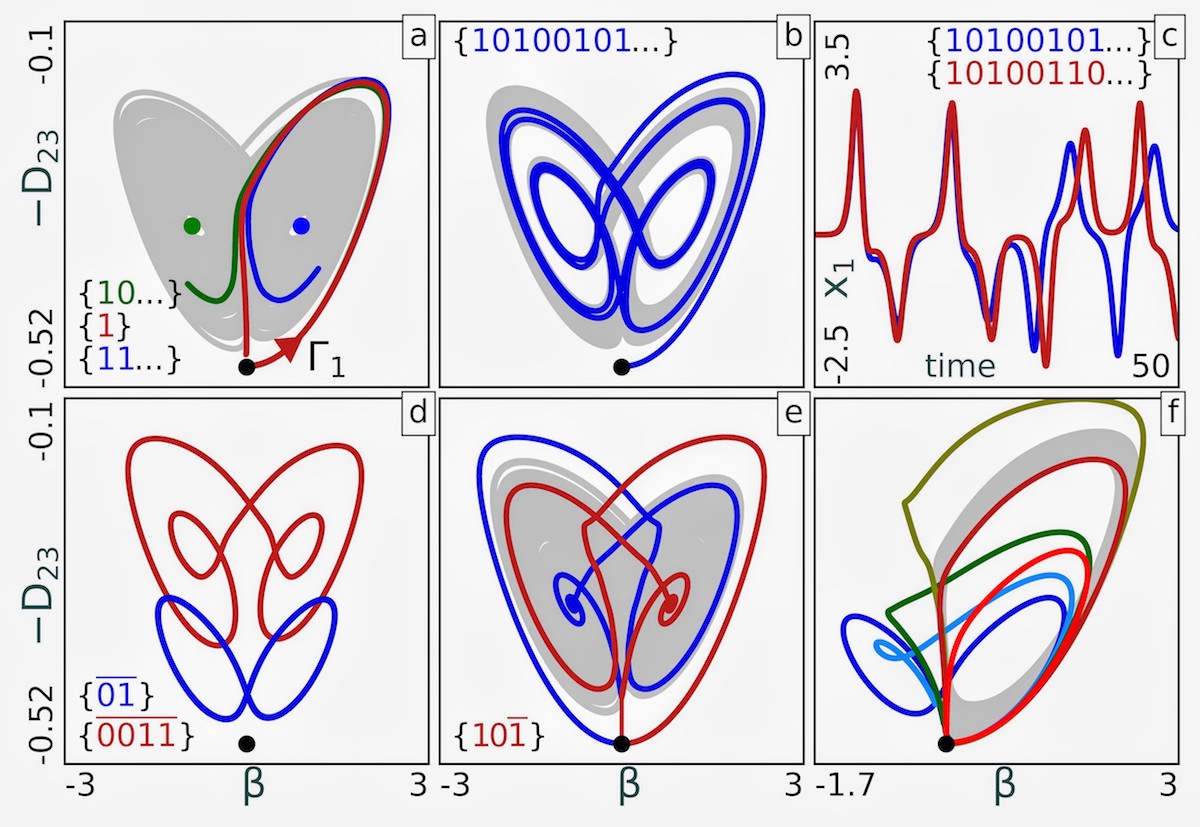}
    \end{center}
    \caption{%
    (color online)
         (a)  ($\beta,-D_{23}$)-phase space projection showing the primary homoclinic orbit (red, coded as $\{1\}$) splitting leftward/rightward (green/blue, $\{11...\}$ or $\{11...\}$)  when the separatrix $\Gamma_1$  misses the saddle $O$ (black dot) after completing a single turn around the saddle-focus $C^+$, with the Lorenz attractor (in grey) in background.  (b)   Chaotic transient of $\Gamma_1$  generating a binary sequence starring with $\{10100101...\}$.  (c) Time-evolutions of the $\beta$-coordinate of $\Gamma_1$ (in (b)) and of a close trajectory (red), and their binary codes, before they diverge. (d) Two stable symmetric POs coded as $\{\overline{01}\}$) and $\{\overline{0011}\}$.  (e)  Heteroclinic connections (red, $\{10\overline{1}\}$) at the $T_1$-point. (f) Samples ($P_j$) of  the primary homoclinic orbit morphing to a double loop after the inclination-flip, $IF_1$, on the curve $H_0$ in the $(a,b)$-parameter plane in Fig.2; here $\sigma = 1.5$.}%
       \label{fig:fig1}
\end{figure}
Both structural and dynamical instability in laser model~(1) is due to an abundance of homoclinic bifurcations ($HB$) of the saddle $O$, whose 1D unstable separatrix $\Gamma_1$ (and $\Gamma_2$) densely fills out the two spatially-symmetric wings of the  butterfly-shaped strange attractor (Fig.~1a,e) \cite{abs}.  As parameters are varied $\Gamma_1$ constantly and unpredictably changes its flip-flop switching patterns within the Lorenz attractor. These patterns change whenever $\Gamma_1$ comes back to $O$ to undergo a homoclinic bifurcation. This observation is the core for the proposed symbolic approach that converts chaotic and periodic patterns of $\Gamma_1$ around equilibria  $C^\pm$ into binary sequences \(\{k_n\}\) as follows: 
\vspace*{-0.2cm}
\[
k_n = \begin{dcases*}
1, & when the separatrix $\Gamma_1$ turns around \(C^+\); \\
0, & when the separatrix $\Gamma_1$ turns around \(C^-\). 
\end{dcases*}
\]
As such, the periodic sequence $\{ 111 \dots \}$. or $\{ \overline 1\}$, corresponds to $\Gamma_1$ converging to the equilibrium state $C^{+}$ or a periodic orbit emerging from though AH-birfurcation, while the sequence $\{ 100 \dots \}$ or $\{ 1 \overline 0\}$ corresponds to $\Gamma_1$ converging to $C^{-}$ and so forth. Wherever small parameter variations do not change $\Gamma_1$-progressions and hence their binary representations, the system exhibits structurally stable dynamics. It is due to the existence of stable equilibria or periodic orbits, such as a symmetric figure-8 periodic orbits (PO) repetitively turning once or twice around \(C^-\) and \(C^+\)  in Fig.~1d. with corresponding binary sequences $\{\overline{01}\}$ and $\{\overline{0011}\}$, resp.  
\begin{figure}[t!]
     \begin{center}
        \includegraphics[width=\columnwidth]{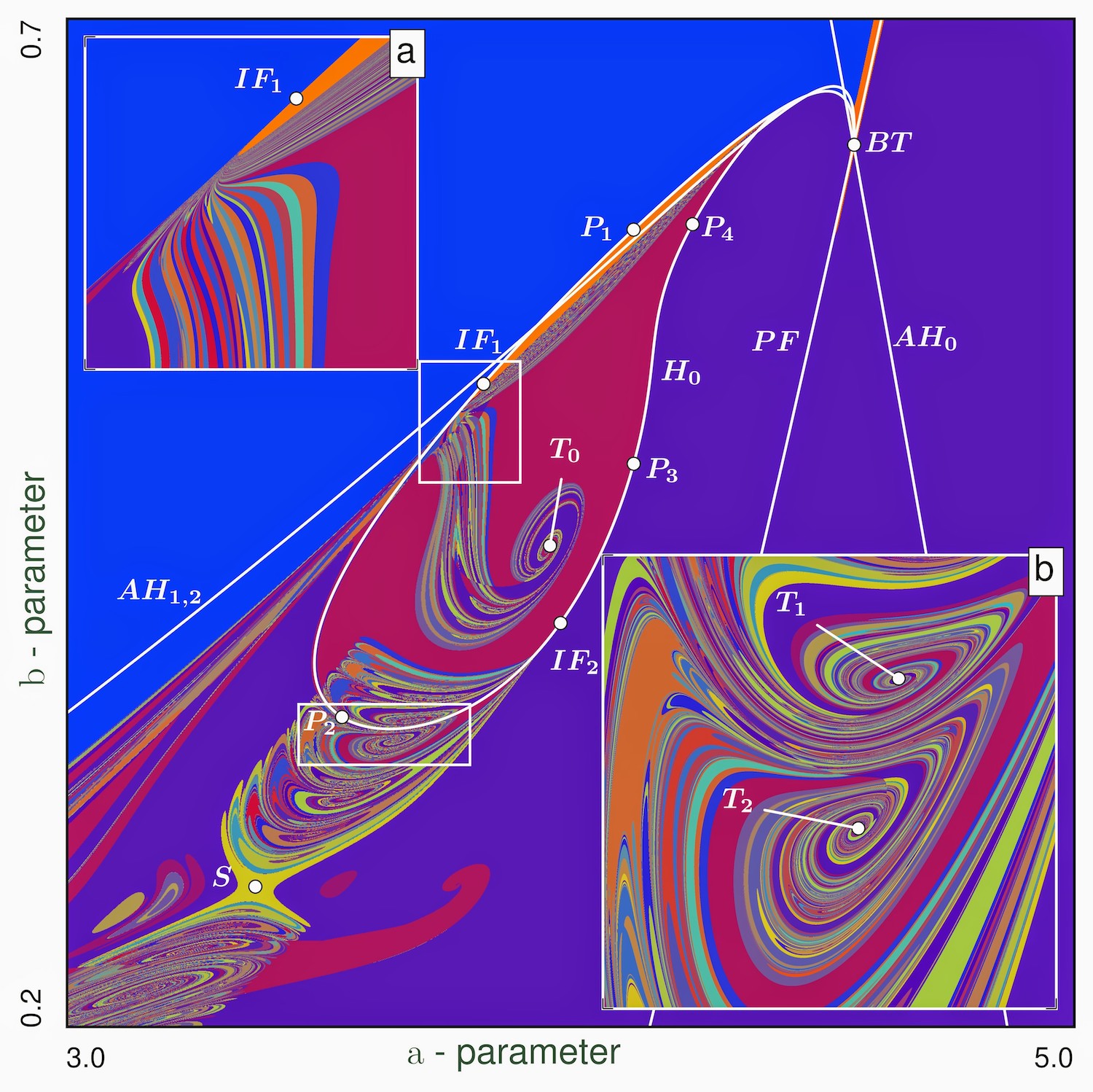}
    \end{center}
    \vspace*{-0.7 cm}
    \caption{%
    (color online)  ($q,\,b$)-parameter sweep of [5--12]-length reveals an abundance of homoclinic bifurcations emerging from two cod-2 points, $IF_1$  \& $IF_2$, on $H_0$, that corresponds to the primary homoclinic butterfly of saddle $0$, along with self-similar characteristic spirals around T-points, labelled $T_{0,1,2}$, corresponding to distinct heteroclinic cycles between $O$ and  saddle-foci $C^\pm$. Cod-2 Bogdanov-Takens, $BT$, unfolding includes Andronov-Hopf $AH_0$, $AH_{1,2}$ and pitch-fork $PF$ bifurcation curves for $O$ and $C^\pm$, resp.; here $\sigma=1.5$. }%
   \label{fig:fig1}
\end{figure}
An aperiodic binary sequence is associated with chaotic dynamics that is characterized by the sensitive 
dependence on small parameter variations that change $\Gamma_1$-progressions and corresponding symbolic 
sequences (Fig.~1c). Changes occurs at homoclinic bifurcations when $\Gamma_1$ comes back to saddle $O$.  
The primary homoclinic orbit (shown in Fig.~1a,f) coded with a finite sequence $\{1\}$ separates periodic patterms coded as $\{ \overline 1\}$ and $\{ 1 \overline 0\}$. It occurs on the bifurcation curve
$H_0$ in the $(a,b)$-parameter plane in Fig.~2. There are two special points labeled as $IF_1$ and 
$IF_2$ on $H_0$ that correspond to the so-called inclination-flip (IF) bifurcation of codimension-two 
\cite{Shilnikov2001}. Its feature is that it gives rise to instant homoclinic chaos in the phase space and to complex bifurcation structures in the parameter space of the system.  With our new computational-symbolic toolkit we can clearly and quickly identify such bifurcations and their fine organizations in the parameter space along with regions of chaotic and regular dynamics.   
\begin{figure}[t!]
     \begin{center}
        \includegraphics[width=\columnwidth]{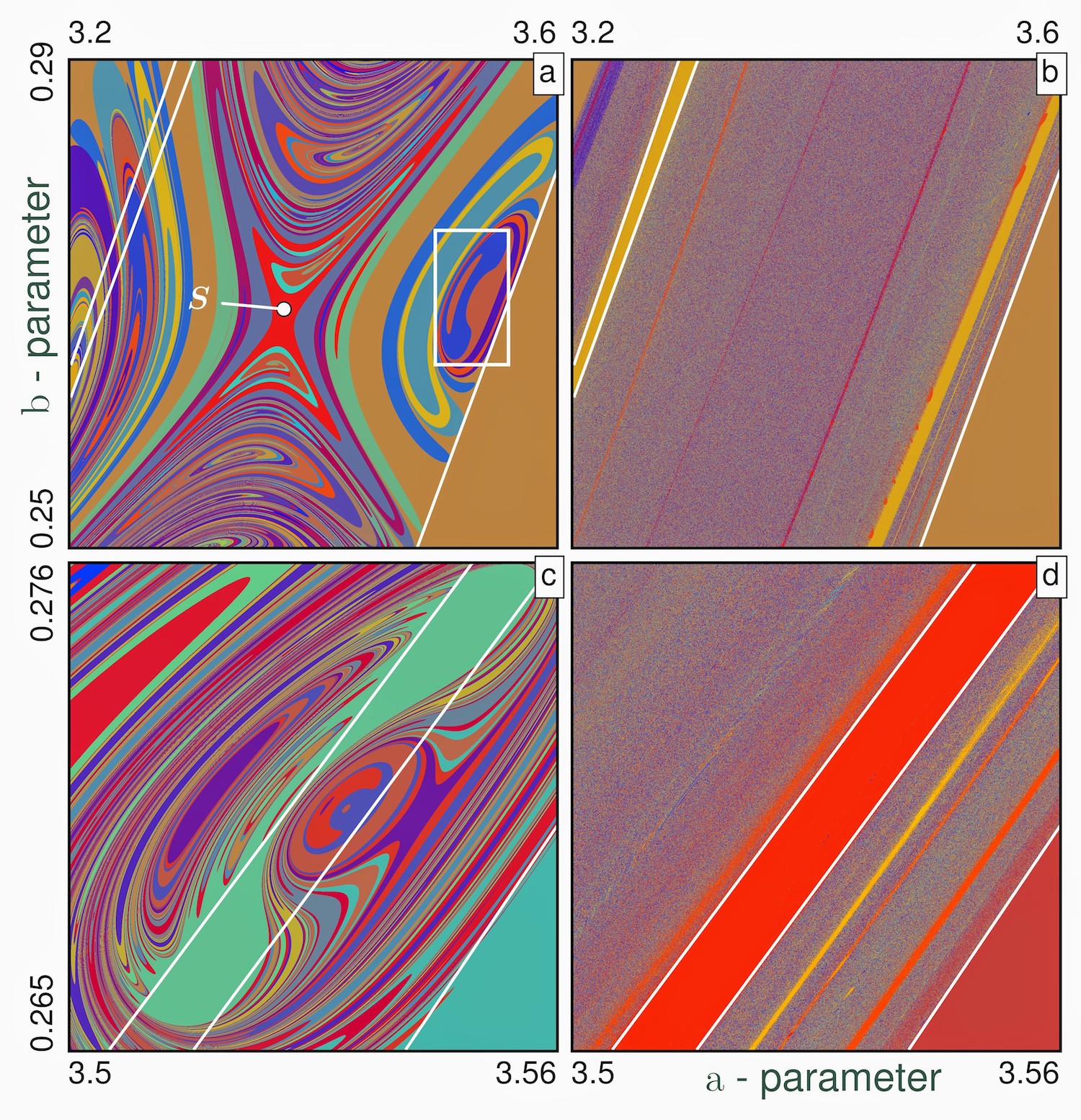}
    \end{center}
    \vspace*{-0.5 cm} 
        \caption{(color online) (a,c) Short [8--15] and (b,d) long [100--123] $(a,b)$-parameter sweeps reveal fine self-similar organizations of homo- and heteroclinic bifurcations underlying the regions of chaotic and regular dynamics of model~(1) for $\sigma = 1.5$. A small area (white box) in  (a) is magnified with a longer [15--22]-sweep in (c). (b,d) reveal stability windows (solid colors) within ``noisy'' regions of structurally unstable chaos; white lines demarcate boundaries of some stability windows.
     }%
   \label{fig:fig1}
\end{figure}
First, we define a formal power series $P(N)$ for a finite binary sequence \(\{k_n\}\) of length $N$, after omitting the first $j$ symbols for initial transients of the separatrix $\Gamma_1$ or any other trajectory, as follows: 
\begin{equation}
P{(N)} = \sum_{n=j+1}^{j+N} \frac{k_n}{2^{(N+j+1)-n}}.
\end{equation}
By construction, the range of $P(N)$ is [0,\,1], including the sequences $\{\overline{0}\}$ and $\{\overline{1}\}$, resp., in the limit as $N \rightarrow \infty$. For example, $P(8)$ for the aperiodic sequence $\{10100101\}$ generated by $\Gamma_1$ in Fig.~1b, with $j=0$ and $N=8$, is given by: 
$P(8) = 1/2^8+0/2^7+1/2^6+0/2^5+0/2^4+1/2^3+0/2^2+1/2^1 = 0.64453125.$ 
\begin{figure}[ht!]
     \begin{center}
        \includegraphics[width=\columnwidth]{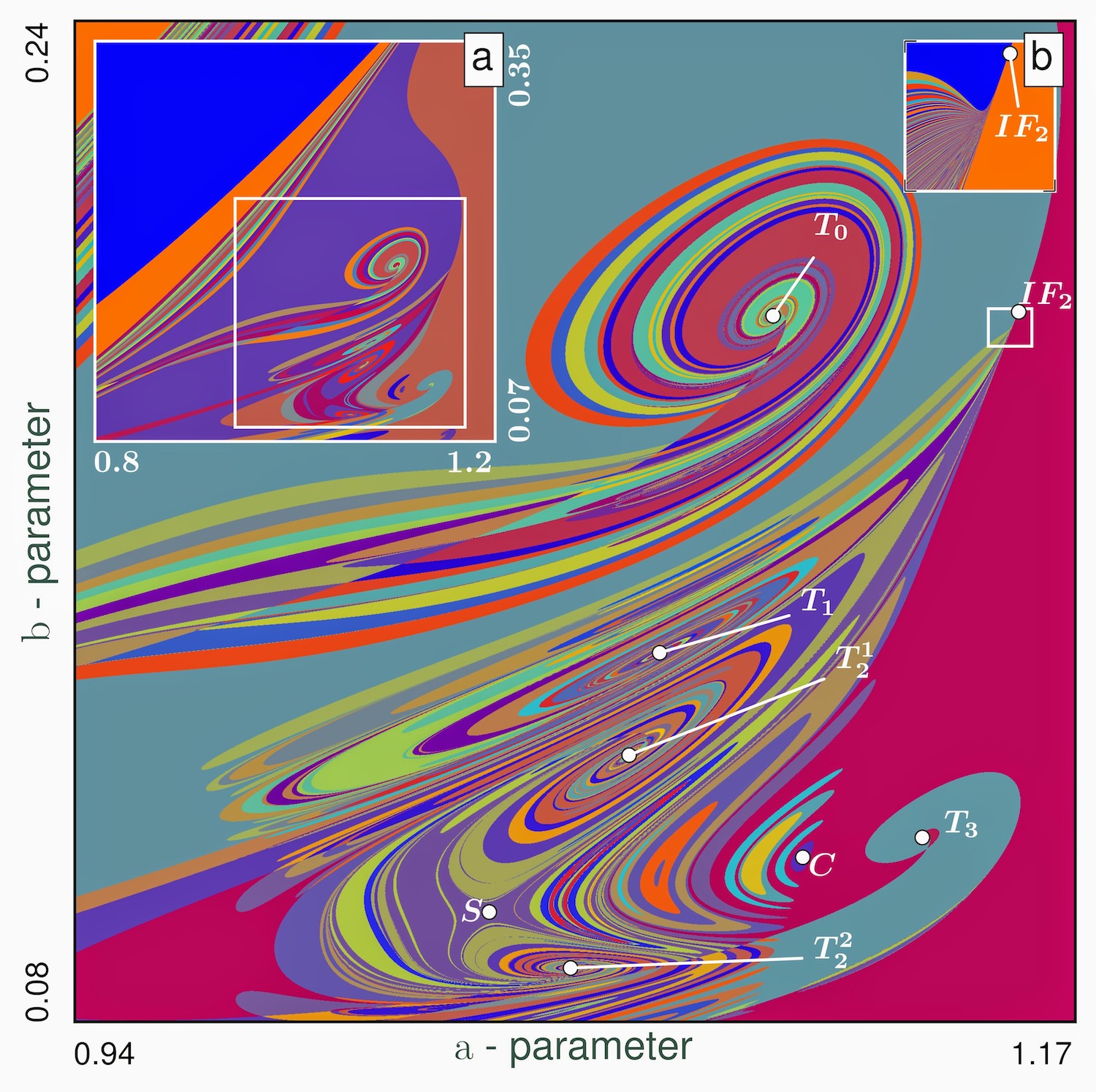}
    \end{center}
    \vspace*{-0.5 cm}
    \caption{%
        (color online)
    [2--9]-length sweep discloses an organization of homo/heteroclinic bifurcations originating from cod-2 inclination-flip $IF_2$ and multiple T-points: primary $T_0$ coded as $\{1\overline{0}\}$, secondary $T_1$ as $\{10\overline{1}\}$, and a pair $T_2^1-T_2^2$ with code $\{11\overline{0}\}$  separated by a saddle (white dot $S$) in the $(a,b)$-parameter plane; here $\sigma=10$.  Inset (a) shows a larger $(a,b)$-sweep of [1--7]-length; (b) [16--23]-long sweep depicts dence loci of homoclinic bifurcation curves originating from $IF_2$.
     }%
   \label{fig:fig1}
\end{figure}
The $P$-quantities are used as invariants to discriminate or conjugate finite progressions of the separatrix $\Gamma_1$ of the saddle against each other to identify and trace down corresponding bifurcation curves in the parameter space. Moreover, the quantities generated from long periodic and aperiodic binary sequences let us efficiently detect regions of regular and chaotic dynamics, resp.    
Keeping $\sigma$ fixed at 1.5 or 10, we 1) vary $a$ and $b$ to a 
bi-parametric sweeps on a 2000x2000 grid 2) to follow $\Gamma_1$-progressions 3) generating binary sequences \(\{k_n\}\) that 4)
result in $P(N)$-quantities. Next 5) we colormap all found $P(N)$ values onto the parameter plane, where
regions are identified by their equivalent colors, and the borderlines between adjacent regions correspond to 
homoclinic bifurcation curves. The colormap differentiates between $P(N)$-values grouped into $2^{24}$ bins 
with pre-preset RGB-color values. Such sweeps can be massively parallelized by running separate threads on a 
graphics processor unit (GPU). For example, the sweep of [5--12]-length, i.e. with first four symbols omitted, 
shown in Fig.~2 takes about 8 seconds to run on Tesla K40 GPU by Nvidia. It is superimposed with the curves, 
obtained by parametric continuation, corresponding to pitch-fork ($PF$), Andronov-Hopf ($AH_0$ and 
$AH_{1,2}$ for $O$ and $C^\pm$) and the primary homoclinic ($H_0$) bifurcations all originating from the 
codimension-2 Bogdanov-Takens point ($BT$) \cite{Shilnikov2001}. Fig.~1f shows how the primary homoclinic 
loop transmutes into a double one along the curve $H_0$.   The sweep reveals the way the inclination-flip 
$IF_1$ and $IF_2$ points give rise to jets of homoclinic bifurcation curves spiraling to various self-similar
cod-2 Bykov terminal T-points, including $T_0$ and $T_1$ corresponds to heteroclinic connections linking the 
saddle $O$ with saddle-foci \(C^+\), \(C^-\) (Fig.~1e) and generating periodic sequences $
\{1\overline{0}\}$, $\{10\overline{1}\}$, resp. 
Figure~3a shows that with longer sequences we can obtain more detailed sweeps disclosing multiple T-points of smaller scales near the saddle point, $S$, that are not seen in Fig.~2. These spiral structures around T-points (identical to $T_2^1$ and $T_2^2$ in Fig.~4) morph into closed loops (like ones shown in Fig.~3c) after collapsing into the saddle through a pitch-fork bifurcation as $\sigma$-parameter is varied (shown in Suppl. Movie~1.) Figures~3b and d present the sweep of [100--123]-length, i.e., after skipping the first 100 transient symbols. Here regions with solid colors of constant $P(23)$-values represent the stability windows corresponding to simple (periodic) Morse-Smale dynamics, whereas multi-colored noisy regions refer to structurally  unstable chaotic dynamics. \\
\noindent
The ($a,\,b$)-sweep of [2-9]-length in Fig.~4 demonstrates  the intrinsic re-arrangement of the bifurcation constituents of complexity for a different cut at $\sigma=10$. Here, the secondary inclination-flip point, ($IF_2$), gives rise to loci of outgoing homoclinic  curves that being re-directed by a saddle point  ($S$), spirals onto multiple T-points. The heteroclinic connections at the T-points, $T_0$-$T_4$, are given by $\{1\overline{0}\}$, $\{10\overline{1}\}$, $\{11\overline{0}\}$, and $\{\overline{1}\}$, respectively. The T-points $T_2^1$ and $T_2^2$, separated by the saddle $S$, correspond to the same heteroclinic connection $\{11\overline{0}\}$. Note that  here the primary homoclinic curve spirals onto the primary T-point $T_1$.  The T-point $T_3$ is located in belongs to the stability window dominated by  the symmetric figure-8 periodic orbit (Fig.~1d) in the long run. The semi-annular structures around $C$ are, in fact, the remnants of the spirals around $T_3$, where the other halves of the spirals are disintegrated by the stable periodic orbit existing near $T_3$. With small $\sigma$-variations 
 $T_3$  crosses over the stability boundary near $C$, so that both ends of the semi-annular structures merge   to complete spirals around $T_3$ (as demonstrated in Suppl. Movie 2.) Meanwhile, T-points $T_2^1$ and $T_2^2$ merge with the saddle $S$ to transform into concentric cycles. 
  \begin{figure}[t!]
     \begin{center}
        \includegraphics[width=\columnwidth]{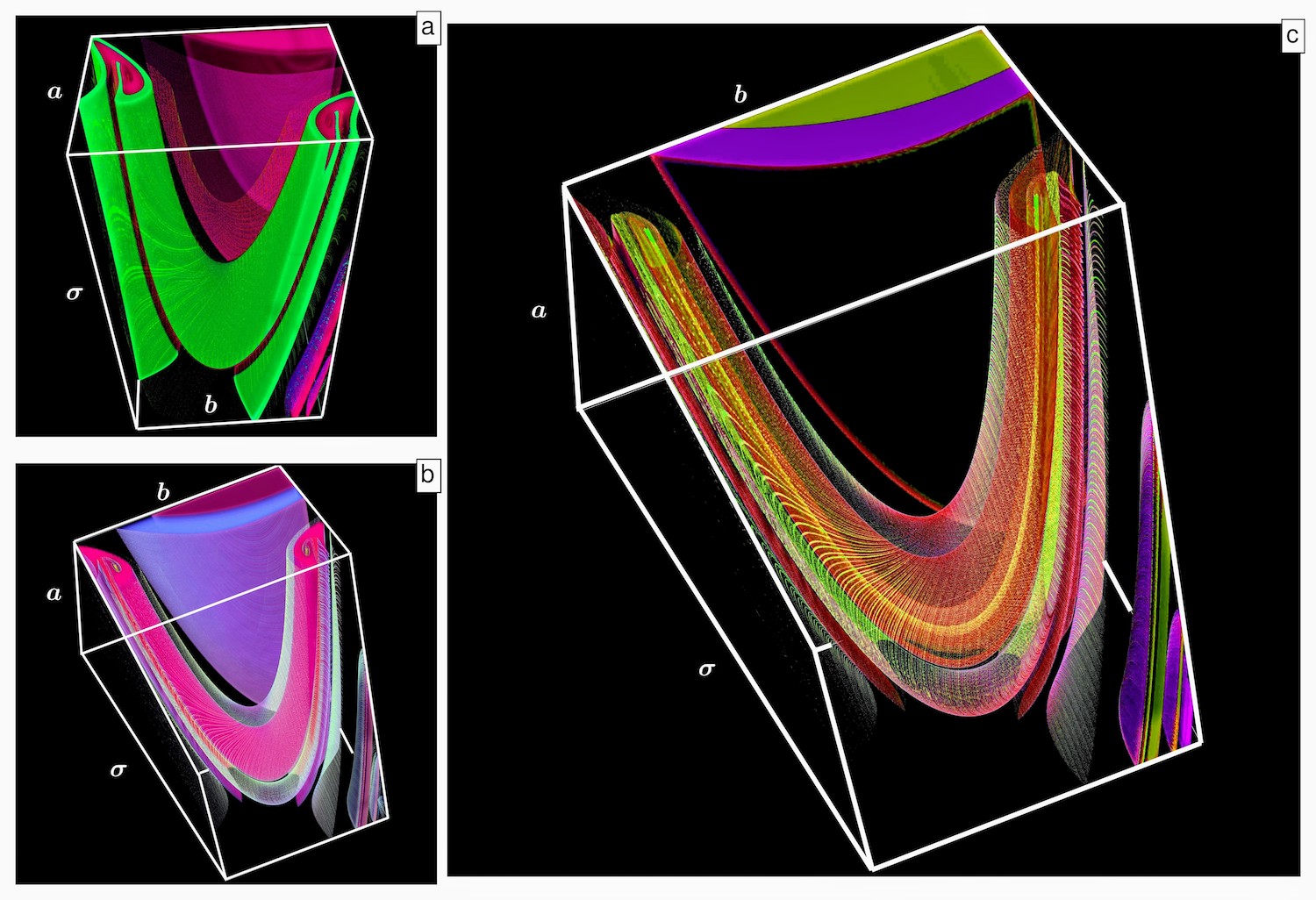}
    \end{center}
    \vspace*{-0.5 cm}
    \caption{%
    (color online)
    Fragment of the 
         3D ($a,b,\sigma$)-parameter space of laser model~(1) depicting compositions of elliptic and hyperbolic paraboloids whose contour curves become spirals around T-points, or concentric circles and saddles in the 2D bi-parametric projections.}
   \label{fig:fig1}
\end{figure}
These structures in the 2D sweeps  are the contour curves of the corresponding surfaces in the 3D  $(a,b,\sigma)$-parameter space of model~(1). Figure~5 shows its near this saddle, which is the critical point of of the 2D surface shaped as a hyperbolic paraboloid. Depending on the particular $\sigma$-cuts the contour lines of the bended scroll-shaped surfaces look like spirals or closed concentric cycles in the projections in Figs.~2-4. \\ 
 \begin{figure}[t!]
     \begin{center}
        \includegraphics[width=\columnwidth]{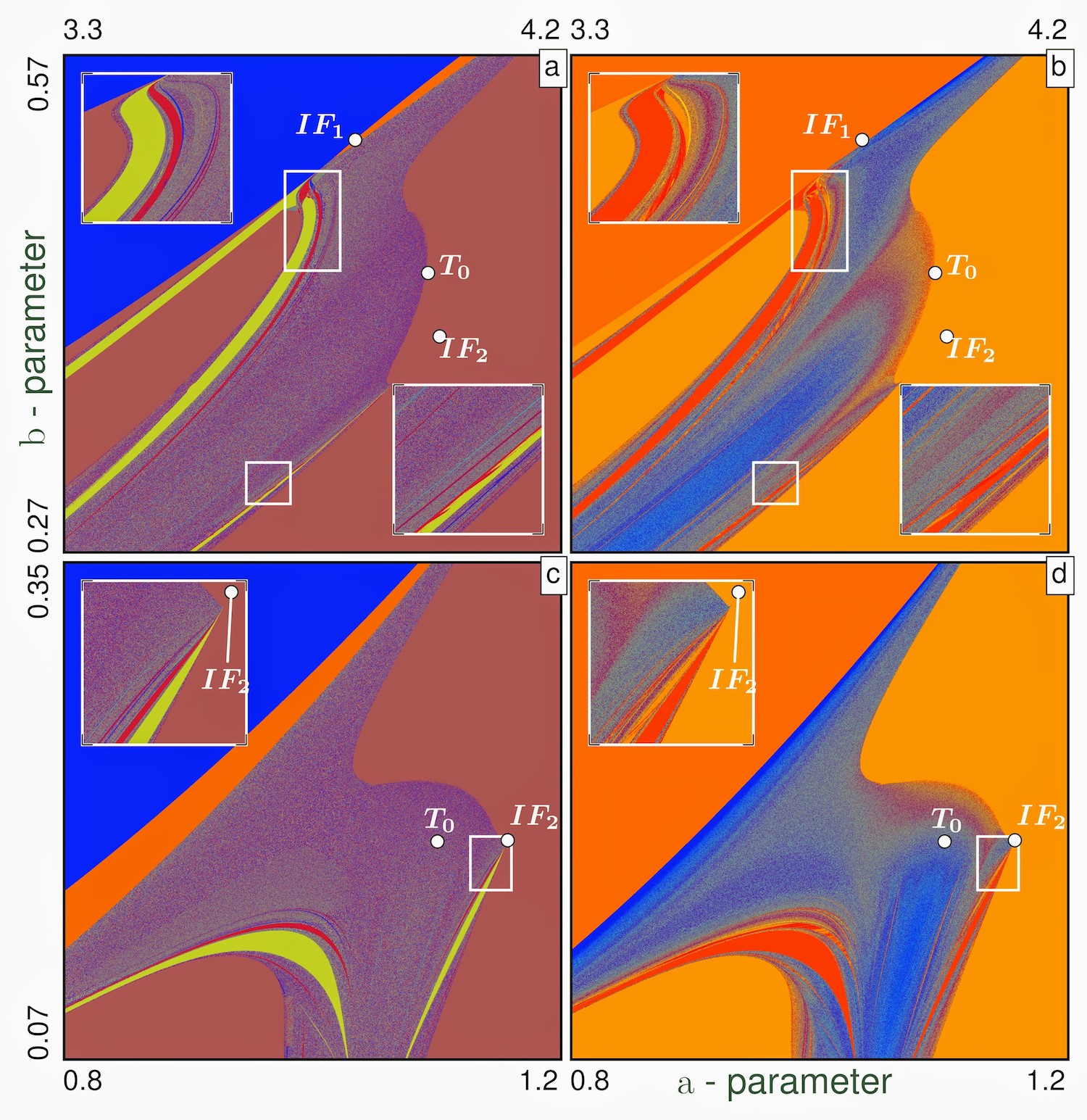}
    \end{center}
    \vspace*{-0.5 cm}
    \caption{%
    (color online) Long [1000--1999]-length sweeps to detect multiple stability windows (solid colors; light green due to  stable PO $\{\overline{0011}\}$) in Fig.~1d) within  (noisy/multi-color) regions of chaos adjacent to  $IF_1$ and $IF_1$ points in the $(a,b)$-parameter space using the proposed DSP symbolic algorithm in (a) and (c), and using LZ-complexity notion in  (b) and (d) $\sigma=1.5$ and $\sigma=10$, resp. to compare with the bifurcation diagram in Fig.~2 and 4.}
   \label{fig:fig1}
\end{figure}
\noindent
While a detailed sweep for short-term transient dynamics lets us reveal the underlying homoclinic bifurcations, longer sweeps, omitting initial transients, are designed to localize stability windows corresponding to regular dynamics [of Morse-Smale systems] and regions of chaotic dynamics in the parameter space. We implemented two algorithms into our computational DSP toolkit to classify such regions depending on whether the corresponding binary sequences of solutions are periodic or not for a given parameter values. The first algorithm based on Eq.~(2) uses additionally a periodicity correction (PC) that identifies the periodic structure within a sequence, and then normalizes it to the smallest valued circular permutation of the periodic sequence. For example, the symmetric figure-8 periodic orbit in Fig.~1d is coded with $\{\overline{01}\}$) not with $\{\overline{10}\}$.  The second algorithm utilizes the Lempel-Ziv-76 (LZ) compression \cite{lz76}, to determine the normalized complexity (the number of words in vocabulary per the sequence length) of the binary sequence. The LZ compression algorithm scans a sequence from left to right, and adds a new word to the vocabulary every time a previously non-encountered substring is detected. Since all circular permutations of a periodic orbit have the same  identical complexity, with this approach we can also detect stability windows amidst structurally unstable chaotic regions. This approach requiring only one solution per a parameter point  complements more expensive computational approaches based on the evaluations of the largest or several  Lyapunov exponents.\\
\noindent
Figure~6 represents the bi-parameter long sweeps of [1000-1999]-length to identify regions where the dynamics of model~(1) is simple and complex, where insets a/c and b/d represent the PC- and LZ-algorithm based sweeps, respectively. Regions of solid monotone colors correspond to the stability windows with stable equilibrium states and periodic orbits, while multi-colored noisy regions indicate that are the dynamics is structurally unstable and chaotic.  The sweeps in Figs.~5a-b (at $\sigma=1.5$)  are superimposed with the primary and secondary inclination-flip points, $IF_1$ and $IF_2$, along with the primary T-point $T_0$ located next to the boundary between the regions of chaotic and stable periodic dynamics.  They reveal multiple stability windows adjacent to $IF_1$ and to $IF_2$ (magnified insets), including the wide one (in light green)  corresponding to a stable periodic orbit $\{\overline{0011}\}$ (shown in Fig.~1d). This approach can  clearly identify distinct periodic orbits and their stability windows mapped by different colors, which  is not possible with the sweeps based on Lyapunov exponents.  Note that identical stability windows (indicated with same colors) emerge near both $IF_1$ and $IF_2$ in the reversed order. The sweeps in Figs.~5c-d (at $\sigma=10$) depict the primary T-point $T_0$ located inside   the region of chaotic dynamics, and stability windows accumulating to $IF_2$. We note that while the PC-algorithm lets one detect and identify a variety of stable periodic orbits efficiently even with short  symbolic sequences (see Figs.~2b,d) compared to quite long sequences required by the LZ-algorithm that  suits better for the detection of chaotic regions. This observation suggests the order to analyze the given sequence and run it first through the PC-algorithm to detect periodic orbits, and next through the LZ-algorithm to detect complexity of aperiodic strings within a minute on Nvidia Tesla K40 GPU.  Other future enhancements for the DSP-toolkit are to include the search algorithms for bifurcations of equilibrium states and periodic orbits such as period-doubling. \\ 
\noindent
In conclusion, we have demonstrated the proficiency of the new symbolic toolkit for computational studies of  both short-term  transient and long-term solutions and to analyze the bifurcation mechanisms underlying  
the onset of chaotic and regular dynamics in the phase and parameter space of the given OPL model and similar deterministic systems.\\  
\noindent
This work was in part funded by NSF grant IOS-1455527, RSF grant 14-41-00044 at the Lobachevsky University of Nizhny Novgorod, RFFI grant 11-01-00001, and MESRF project 14.740.11.0919. We thank GSU Brain and Behaviors Initiative for the fellowship and pilot grant support and are grateful to all members of Shilnikov's NeurDS lab for the helpful discussions. 


%

\end{document}